\documentclass[%
 reprint,
 amsmath,amssymb]{revtex4-2}

\usepackage{CJK}
\usepackage{graphicx}% Include figure files
\usepackage{dcolumn}% Align table columns on decimal point
\usepackage{bm}% bold math
\usepackage{float}
\usepackage[dvipsnames]{xcolor}
\usepackage{appendix}
\usepackage{enumerate}% http://ctan.org/pkg/enumerate

\newcommand{\bfq}{\mathbf{q}}
\newcommand{\kB}{k_\mathrm{B}}

\begin{document}

%\preprint{APS/123-QED}
\begin{CJK*}{}{} % Use default fonts from CJK (see below)
\title{Normal liquid $^3$He studied by Path Integral Monte Carlo with a parametrized partition function}
\author{Tommaso Morresi$^{1\ast}$, Giovanni Garberoglio$^{1\ast}$\\~\\}
\affiliation{$^{1}$ European Centre for Theoretical Studies in Nuclear Physics and Related Areas (ECT*), Fondazione Bruno Kessler, Italy}

\begin{abstract}
We compute the energy per particle of normal liquid ${}^3$He in the temperature range $0.15-2$~K using Path Integral Monte Carlo simulations, leveraging a recently proposed method to overcome the sign problem -- a long-standing challenge in many-body fermionic simulations. This approach is based on introducing a parameter $\xi$ into the partition function, which allows a generalization from bosons ($\xi=1$) to fermions ($\xi=-1$). By simulating systems with $\xi \geq 0$, where the sign problem is absent, one can then extrapolate to the fermionic case at $\xi = -1$.
Guided by an independent particle model that uncovers non-analytic behavior due to the superfluid transition, which is moderated by finite-size effects, we develop a tailored extrapolation strategy for liquid ${}^3$He that departs from the extrapolation schemes shown to be accurate in those cases were quantum degeneracy effects are weak, and enables accurate results in the presence of Bose--Einstein Condensation and superfluidity for $\xi > 0$. Our approach extends the previously proposed framework and yields energy per particle values in good agreement with experimental data.
\end{abstract}

\maketitle
\end{CJK*}
%\tableofcontents

\section{Introduction}
Computer simulations of interacting many-fermion systems are notoriously difficult due to the 
requirement of antisymmetric quantum states. If one aims at exact solutions (that is, avoiding uncontrolled approximations), numerical methods such as the Full Configuration Interaction scale exponentially with the system size,~\cite{Szabo12} preventing calculations to be performed except for the smallest systems. In the case of electronic structure calculations, this currently limits the size of the systems that can be studied to tens of electrons. For larger systems, approximations such as the Coupled Cluster approach provide results that are "chemically accurate" when single, double and perturbative triplet excitations are considered, at the cost of scaling with the number of electrons $N$ as ${\cal O}(N^7)$.~\cite{Bartlett07}

Monte Carlo methods are generally expected to provide a better scaling, but in this case the antisymmetry requirement results in the so called {\em sign problem}; sampling regions of positive and negative values of the density matrix results in a large variance of the various observables,~\cite{loh_1990,Lipparini08,dornheim_2019} again limiting the size of the systems that can be studied. Recently, an extensive Path Integral Monte Carlo (PIMC) simulation of liquid ${}^3$He could not exceed $N=38$ even when resorting to large-scale computational resources.~\cite{dornheim_he3}
Several approximations, such as the famous "fixed nodes" one,~\cite{anderson_1975,ceperley_1992,turborvb} provide good quality results with a favorable scaling of ${\cal O}(N^3)$ with the number of particles.~\cite{Lipparini08}

In the past few years, a novel method to overcome the sign problem for many-body fermionic simulations has been introduced. This approach is based on the consideration that the quantum statistical partition function of a system of bosons and fermions differs only by the value a single parameter, $\xi$, which assumes the value $\xi=1$ for bosons and $\xi=-1$ for fermions.~\cite{dornheim_2023_2,xiong_2022}
Since the partition function turns out to be a {\em continuous} function of $\xi$, one can envision an extrapolation procedure to the fermion case, based on results in the $\xi \geq 0$ regime, where PIMC calculations are not affected by the sign problem and can be carried out with acceptable computational resources.
This approach has been successfully applied to the study of the electron gas~\cite{dornheim_2023_2} and solvable fermion models.~\cite{xiong_2023} In practice, one can directly extrapolate the energy (or other observables such as the pair correlation functions) obtained at different $\xi\geq0$. But in a more general framework,~\cite{xiong_2023} that will also be used in this work, one can first compute the average energy per particle $E$ (or, in principle, also other observables) along an isotherm of temperature $T$ for several values of $\xi \geq 0$. These isotherms are  then used to compute how $\xi$ depends on $T$ for fixed energy $E$. It has been shown that simple fitting functions, usually quadratic polynomials, for $\xi_E(T)$ extrapolate very well to $\xi=-1$, thus providing the temperature $T$ at which the fermionic system attains energy $E$.~\cite{xiong_2023} 
However, the application of this extrapolation method so far has been performed under condition of weak quantum degeneracy effect.~\cite{dornheim_2023_2}

In this work, we investigate how this approach can be used to compute $E(T)$ in a case of normal liquid ${}^3$He, where quantum degeneracy is significant. We find that the presence of Bose--Einstein Condensation (BEC) and the accompanying superfluid transition in the $\xi > 0$ region results in a non-analytic behavior for the function $\xi_E(T)$ which, despite being smoothed out by finite-size effects, requires more care in extrapolating to $\xi = -1$  than in the case of previous studies.~\cite{xiong_2023} %electrons.
After recalling the general theoretical framework, we will discuss an independent particle model that highlights the effect of BEC on the values of $E(T,\xi)$ along an isotherm and provides important clues on the general form of $\xi_E(T)$, %in this case, 
that hold for both the fully spin-polarized and spin-unpolarized cases. Secondly, we will study finite-size effects in the same model, and investigate how they affect the extrapolation procedure. Finally, based on extensive PIMC simulations of $E(T, \xi)$, we apply our extrapolation method to compute $E(T)$ for ${}^3$He, finding a %very 
good agreement with experimental data and previous theoretical results.

\section{Theoretical framework}
The quantum statistical partition function for $N$ identical particles at temperature $T$ can be written as
\begin{equation}
%    Z(\beta) = \text{Tr}\left[ e^{-\beta \hat{H}}\right],
    Z_{\pm}(T) = \frac{1}{N!}\sum_{\mathcal{P}} (\pm 1)^{\mathcal{P}} \int d \mathbf{X} \ \rho \left( \mathbf{X},\mathcal{P} \mathbf{X},\beta \right),
    \label{eq:z}
\end{equation}
where $\beta = (\kB T)^{-1}$, $\kB$ is the Boltzmann constant, the sign '$+$' is for bosons and '$-$' for fermions, and $\mathbf{X}=\left(\mathbf{r}_1, s_1; \mathbf{r}_2, s_2; \dots,\mathbf{r}_N, s_N\right)$ is a compact notation for all the particle coordinates, that is positions $\mathbf{r}_i$ and spins $s_i$.
In Eq.~(\ref{eq:z}), $\rho(\mathbf{X},\mathbf{X}',\beta)=\langle \mathbf{X}|e^{-\beta H} | \mathbf{X}' \rangle$ is the canonical density matrix corresponding to the Hamiltonian $H$, and $\mathcal{P} \mathbf{X}=(\mathbf{r}_{\mathcal{P}(1)}, \sigma_{\mathcal{P}(1)}; \mathbf{r}_{\mathcal{P}(2)}, \sigma_{\mathcal{P}(2)}; \dots,\mathbf{r}_{\mathcal{P}(N)}, \sigma_{\mathcal{P}(N)})$ denotes the position vectors with permuted labels.

One can generalize Eq.~(\ref{eq:z}) introducing a real parameter $\xi$ and writing %a parametrized partition function, which reads as
\begin{equation}\label{eq:zxi}
    Z_\xi(T) %\sim 
    =\frac{1}{N!}\sum_{\mathcal{P}} \xi^{N_{\mathcal{P}}} \int d \mathbf{X} \ \rho \left( \mathbf{X},\mathcal{P} \mathbf{X},\beta \right).
\end{equation} 
where $N_{\mathcal{P}}$ denotes the minimum number of times for which pairs of indices have to be interchanged in the current permutation $\mathcal{P}$ to recover the original (diagonal) order.~\cite{xiong_2022,xiong_2023,dornheim_2023_2,dornheim_2024,dornheim_2024_1}
In the following, we will be concerned with a spin-independent Hamiltonian $H$, hence it is convenient to write the states in the factorized form $|\mathbf{X}\rangle = |\mathbf{R}\rangle |\mathbf{S}\rangle$, where $\mathbf{R}$ is a shorthand for all the positions $\mathbf{r}_i$ and $S$ denote all the single-particle spins; the density matrix can then be written
\begin{equation}
\rho(\mathbf{X}, {\cal P} \mathbf{X}, \beta) = 
\rho(\mathbf{R}, {\cal P} \mathbf{R}, \beta)
\langle S | {\cal P} S \rangle.
\label{eq:rho_factorized}
\end{equation}

Equation~(\ref{eq:rho_factorized}) shows that the density matrix $\rho$ is zero, unless the permutation operator ${\cal P}$ is such that it does not exchange particles with different spins. Considering particles with spin $\frac{1}{2}$, this requirement implies ${\cal P} = {\cal P}_- {\cal P}_+$, where ${\cal P}_-$ is a permutation among particles with negative spin, and ${\cal P}_+$ is a permutation among particles with positive spin. This is equivalent to saying that since the single-particle spin is a good quantum number, particles of different spins are distinguishable~\cite{Zong98} and the total spin $S$ is constant.
This property enables simulations to be performed at different polarizations, by randomly assigning a fraction $\phi$ of the particles to positive spin, and -- correspondingly -- a fraction $1-\phi$ of the particles to negative spin, resulting in polarization %$\zeta \equiv S/N = 2\phi - 1$. 
$P\equiv S/N = 2\phi - 1$
During the Monte Carlo calculation, one independently samples the permutation among particles of positive or negative spin.

In Eq.~(\ref{eq:zxi}), the parameter $\xi$ continuously interpolates from bosons ($\xi$ = 1), to distinguishable particles ($\xi = 0$), to fermions ($\xi =-1$), and can take any real value in ($-\infty$,+$\infty$).
The energy of the system described by Eq.~(\ref{eq:zxi}) can be obtained from standard statistical mechanics considerations as
\begin{equation}
    E_{\xi}(T) = -\frac{1}{Z_{\xi}(T)}\frac{\partial Z_{\xi} (T)}{\partial \beta},
\end{equation}
where the subscript means the $\xi$ variable is kept fixed. 
In order to estimate the energy in the case of fermionic systems, one can then exploit the fact that both $Z_\xi(T)$ and $E_\xi(T)$ are continuous  functions of $T$ and $\xi$. By knowing their behavior in the $\xi \geq 0$ sector, where calculations can be performed without encountering the sign problem, one can extrapolate the value of the energy to $\xi = -1$.
In fact, for $\xi\ge 0$ one can perform standard PIMC simulations~\cite{ceperley_1995} just by taking into account the different positive weights due to the factor $\xi^{N_{\mathcal{P}}}$ in Eq.~(\ref{eq:zxi}). Notice that in the "bosonic" ($\xi > 0$) regime, one is in general simulating a system that is a mixture of different bosons, one for each of the $2s+1$ values of the spin of the limiting fermionic particles. Only in the case of a fully polarized system ($P = 1$) does the bosonic sector correspond to actual pure bosons.

In the case of a spin-independent Hamiltonian, there are two necessary conditions that can be used to guide the extrapolation to $\xi = -1$:~\cite{xiong_2022,xiong_2024}
\begin{enumerate}
    \item $\frac{\partial E_{T}(\xi)}{\partial \xi} < 0$. The parameter $\xi$ induces an effectively repulsive exchange interaction for fermions and effectively attractive exchange interaction for bosons; thus for a given fixed temperature the energy should decrease with increasing $\xi$.
    \item $\frac{\partial E_{\xi}(T)}{\partial T} = -\frac{1}{T^2} \frac{\partial E_{\xi}(T)}{\partial \beta} > 0$. This is the condition to have a positive heat capacity.
\end{enumerate} 

In a first paper~\cite{xiong_2022}, the authors proposed a parabolic behavior of the energy with respect to $\xi$:
\begin{equation}
    E(T,\xi) = c_0 + c_1 \xi + c_2 \xi^2, 
    \label{eq:E_extrap}
\end{equation}
for each temperature. They found that this empirical formula worked quite well for different repulsive interactions, such as the Coulomb, the dipolar and the Gaussian ones.~\cite{xiong_2022} The same approach has been shown to give a good description of large Fermi-systems of weak quantum degeneracy, but it was found to break down for moderate to high quantum degeneracy.~\cite{dornheim_2023_2} Also, the parabolic extrapolation has been used for other observables such as the static structure factor. Recently, the extrapolation approach has been successfully adopted to describe warm dense hydrogen and beryllium.~\cite{dornheim_2024}

A more general approach was then proposed in Ref.~\cite{xiong_2023}. Based on the same two physical assumptions previously mentioned, the authors generalised the theory by considering $\xi$ as a function of both $T$ and $E$. 
In particular, they proved the following relation:
\begin{equation}\label{eq:dcsi_dt}
    \frac{\partial \xi_{E}(T)}{\partial T} \Bigg|_{T=0} = 0,
\end{equation}
resulting in the lack of a the linear term when expanding $\xi$ as a function of $T$, {\em i.e.},
\begin{equation}\label{eq:csi_exp_xiong}
    \xi_E(T) = a_0(E) + a_2(E) T^2 + \sum_{i>2} a_i(E) T^i.
\end{equation}
In practice, one computes numerically $E(\xi)$ for several values of $T$, samples the function $\xi_E(T)$ and fits the parameters $a_i$ using Eq.~(\ref{eq:csi_exp_xiong}). Finally, the temperature $T_E$ where the fermionic system attains energy $E$ is obtained by solving 
$\xi_E(T_E)=-1$. In their analysis with free electrons, keeping only the quadratic term in Eq.~(\ref{eq:csi_exp_xiong}) was enough to provide accurate results.~\cite{xiong_2023} 

While the extrapolation approach of Eq.~(\ref{eq:E_extrap}) has already been used to study observables different from the energy per particle ({\em e.g.}, the pair correlation function,~\cite{xiong_2022}), we stress that the possibility of extending the present approach based on the extrapolation of Eq.~(\ref{eq:csi_exp_xiong})~\cite{xiong_2023} to study other observables is feasible and will be investigated in future studies.

\section{Independent-particle model}
Before presenting the PIMC results for $^3$He, let us investigate an independent-particle model system. In this case, the relation between $\xi$ and $T$ must be computed exactly, but some analytic considerations highlight the effect of BEC. 
The generalized quantum statistical mechanics as a function of $\xi$, that is the statistical mechanics of particles obeying the generalized commutation relation
\begin{equation}
    a_\bfq a^\dagger_{\bfq'} - \xi a^\dagger_{\bfq'} a_\bfq = \delta_{\bfq, {\bfq'}},
\end{equation}
has been studied in Ref.~\onlinecite{Isakov93}, where it is shown that in an independent-particle model having Hamiltonian $H = \sum_\bfq \epsilon(\bfq) \hat{N}_\bfq$ (where $\hat{N}_\bfq$ is the usual number operator), occupation numbers have the form
\begin{equation}
    f(\epsilon; T, \mu) = \frac{1}{\exp\left(\frac{\epsilon-\mu}{T}\right) - \xi},
    \label{eq:occupation_numbers}
\end{equation}
where $\mu$ is the chemical potential.
In the grand canonical ensemble, one obtains the following coupled equation relating the temperature $T$ and the chemical potential $\mu$ to the number of particles $N$ and the average energy $E$ 
\begin{align}\label{eq:gc_N}
        N &= N_\mathrm{c} + \frac{\Omega}{(2\pi)^3} \int_0^{\infty} \frac{4 \pi q^2}{\text{exp}\left( \frac{\epsilon(q)-\mu}{T} \right) -\xi}\mathrm{d}q \\
        \label{eq:gc_E}
        E &= \frac{\Omega}{(2\pi)^3} \int_0^{\infty} \frac{4 \pi q^2 \epsilon(q)}{\text{exp}\left( \frac{\epsilon(q)-\mu}{T} \right) -\xi}\mathrm{d}q,
\end{align}
where $\Omega$ is the volume of the system and the dispersion relation depends only on $q = |\bfq|$ due to spherical symmetry. We will assume in the following that $\epsilon(q > 0) > 0$ and that $\epsilon(q\to0) = 0$. The symbol $N_\mathrm{c}$ in Eq.~(\ref{eq:gc_N}) represents the non-zero macroscopic number of particles in the ground state ({\em i.e.}, the state with $\epsilon = 0$) that signals the onset of BEC, which is characterized by the fact that $N_\mathrm{c}/N$ remains finite in the thermodynamic limit.~\cite{Pitaevskii16} The presence of BEC in this system is related to the fact that for any $\xi > 0$ the chemical potential has an upper bound, lest the occupation numbers of Eq.~(\ref{eq:occupation_numbers}) become negative. Inspection of Eq.~(\ref{eq:occupation_numbers}) shows that the chemical potential must be smaller than the critical value $\mu_\mathrm{c} = - T \log\xi$.
Conversely, if one considers Eq.~(\ref{eq:gc_N}) as a function of $\xi$ along an isotherm, there is a critical value of the parameter $\xi$ after which $N_\mathrm{c} > 0$ given by
\begin{align}\label{eq:xicrit}   
\xi_{\mathrm{c}} &= \frac{\Omega}{N (2\pi)^3} \int_0^{\infty} \frac{4 \pi q^2}{\text{exp}\left( \frac{\epsilon(q)}{T} \right) -1}\mathrm{d}q.
\end{align}
In correspondence to this critical value, the energy as a function of $\xi$ exhibits non-analytic behavior due to the increasing occupation of $N_\mathrm{c}$. The precise form of this non-analyticity depends on the structure of the dispersion relation $\epsilon(q)$ and takes the form
\begin{align}
    \frac{N_\mathrm{c}}{N} \sim 1 - \left( \frac{T}{T_\mathrm{c}} \right)^\alpha,
\end{align}
where $T_\mathrm{c}$ is the BEC transition temperature, while $\alpha=3/2$ for a quadratic dispersion and $\alpha=3$ for a linear dispersion. In the case of ${}^3$He, the dispersion relation is closer to the linear case, which corresponds to a stronger non-analytic behavior.
\begin{figure}
   \includegraphics[scale=0.7]{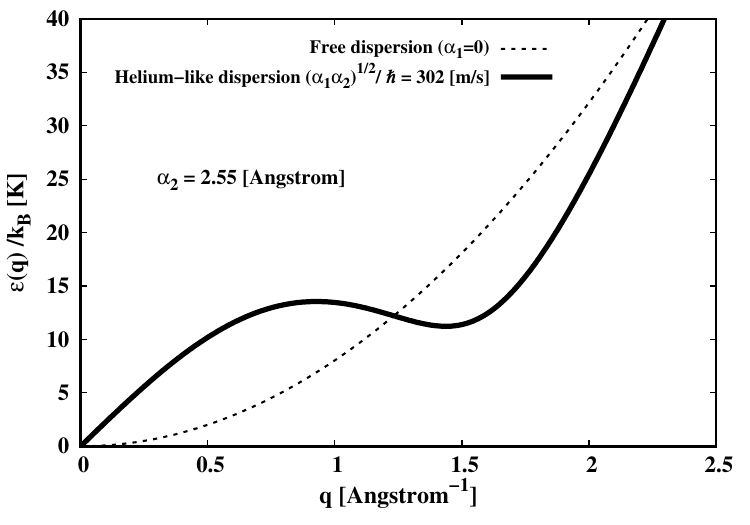}
   \caption{Plot of the single-particle dispersions of Eq.~(\ref{eq:disp_helium-like}) for two different $\alpha_1$ parameters. The dispersion studied in this work corresponds to the continuous line.}\label{fig:disp_models}
\end{figure}

\begin{figure*}
   \includegraphics[scale=0.54]{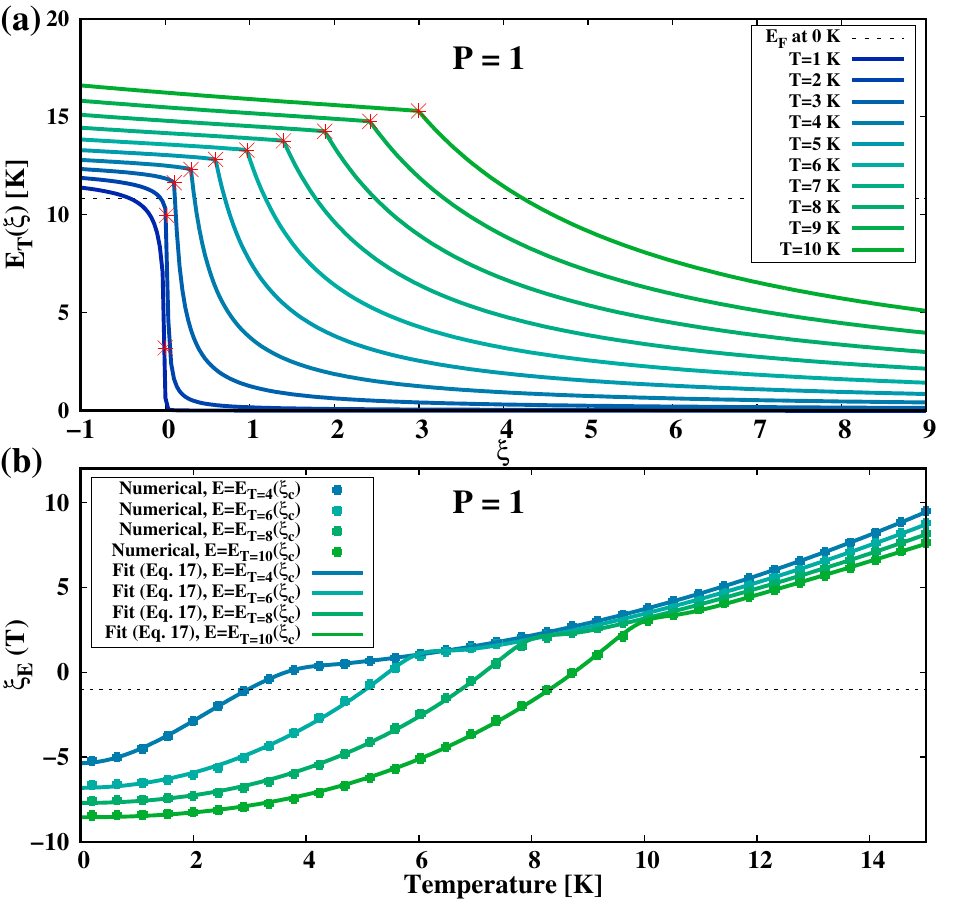}
   \includegraphics[scale=0.54]{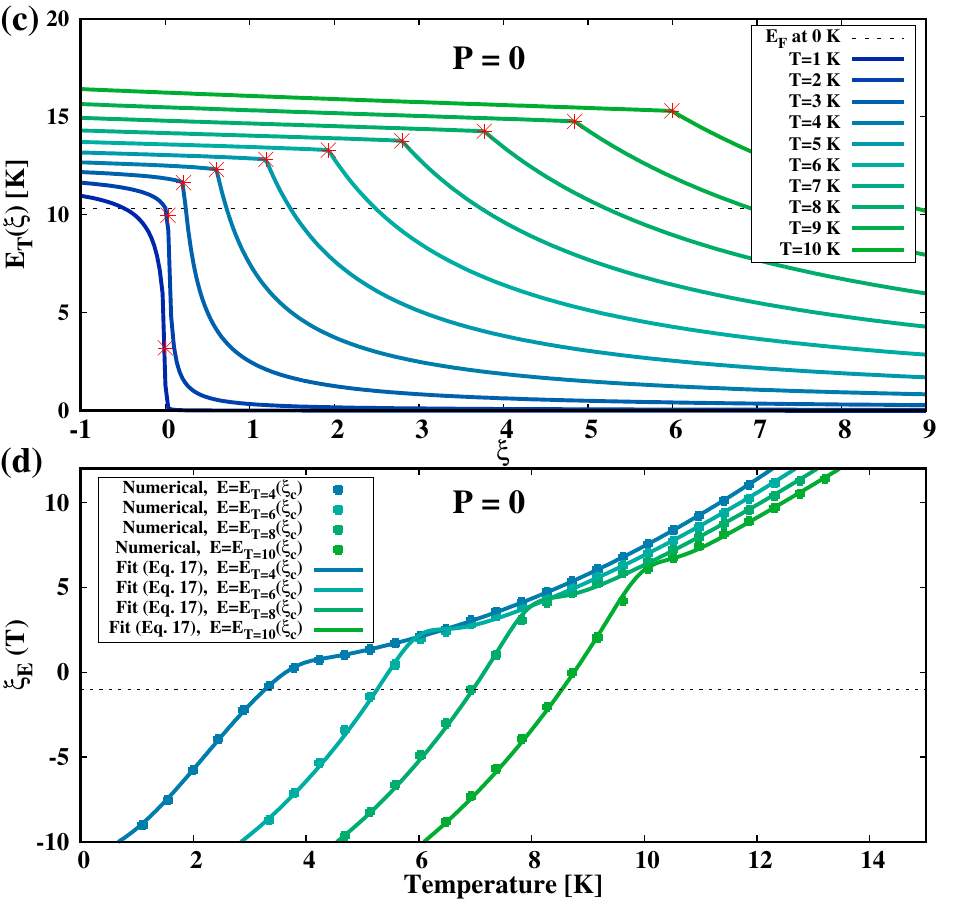}
   \caption{Independent--particle model. (a) Energies as a function of the $\xi$ parameter for the dispersion in Eq.~(\ref{eq:disp_helium-like}) and for $P=1$ (fully polarized case) from Eqs.~(\ref{eq:gc_N}) and (\ref{eq:gc_E}) for temperatures between $1$~K and $10$~K. Red stars indicate the energies obtained at $\xi_{\mathrm{c}}$ in Eq.~(\ref{eq:xicrit}). (b) Behaviour of $\xi_E(T)$ for a few selected energies for $P=1$. The latter are chosen at the critical points for $T=4, 6, 8$, and $10$~K. Notice that the slope change of $\xi_E(T)$ is located at these temperatures. (c) and (d): Same of (a) and (b) in the case of $P=0$ (fully unpolarized system).} 
   \label{fig:res_helike}
\end{figure*}

Equations~(\ref{eq:gc_N}) and (\ref{eq:gc_E}) have been presented, for the sake of simplicity, neglecting the spin variable. In the case of spin-independent Hamiltonians, the model can be straightforwardly generalized since, as discussed before, the $z$ component of the spin is a conserved quantum number. Therefore, in a single-particle approach we can consider our system as a mixture of particles of different spins; each spin component will be described by a pair of equations analogous to (\ref{eq:gc_N}) and (\ref{eq:gc_E}). Specializing to the case of ${}^3$He, we will then have a fixed number of particles with spin up $N_\uparrow$ and a fixed number of particles with spin down $N_\downarrow$ such that $N=N_\uparrow+N_\downarrow$, with corresponding chemical potentials $\mu_\uparrow$ and $\mu_\downarrow$, as well as energies $E_\uparrow$ and $E_\downarrow$. The polarization $P$ of the system will be then constant and equal to
\begin{equation}
    P = \frac{N_\uparrow - N_\downarrow}{N_\uparrow + N_\downarrow},
\end{equation}
and the total energy per particle, $E/N$, will be given by
\begin{equation}
    \frac{E}{N} = \frac{N_\uparrow}{N} E_{\uparrow} + 
     \frac{N_\downarrow}{N} E_{\downarrow}.
\end{equation}
Notice that, according to Eq.~(\ref{eq:xicrit}), the various spin components will undergo BEC separately, that is the function $E_T(\xi)$ will display two non-analytic points, unless the system is unpolarized ($P=0$, hence $\xi_{\mathrm{c},\uparrow} = \xi_{\mathrm{c},\downarrow}$) or fully polarized ($P = -1$ or $P=1$).

In order to solve Eqs.~(\ref{eq:gc_N}) and (\ref{eq:gc_E}), we use a dispersion relation parametrized to mimic the single-particle excitation spectrum of helium~\cite{lemeshko_2017}, which has the form
\begin{equation}\label{eq:disp_helium-like}
    \epsilon (q) =  \sqrt{\alpha_1 q \cdot \text{sin}\left(\alpha_2  q\right) + \left( \frac{\hbar^2 q^2}{2m}\right)^2},
\end{equation}
where $m=3.016$ a.m.u. is the mass of $^3$He.
We notice that by tuning the two parameters, and $\alpha_1$ in particular, one can also obtain a free dispersion ($\alpha_1$=0, short dashed line in Fig.~\ref{fig:disp_models}) or a Bogoliubov-like dispersion ($\sqrt{\alpha_1\alpha_2/\hbar^2} \sim 160 $ [m/s]). Here the two parameters $\alpha_1$ and $\alpha_2$ are taken such that $\alpha_2=2.55 $ \AA \ and $\sqrt{\alpha_1\alpha_2/\hbar^2} = 302 $ [m/s]. The studied dispersion is shown in Fig.~\ref{fig:disp_models} by a continuous line. 
We solved numerically Eqs.~(\ref{eq:gc_N}) and (\ref{eq:gc_E}) fixing the density at $m N/\Omega=0.081$~g/cm$^3$, and report our results in Fig.~\ref{fig:res_helike} for the fully polarized (panels (a) and (b)) and unpolarized (panels (c) and (d)) systems.
Notice the presence of a cusp in the curves $E_T(\xi)$ in correspondence of the critical value $\xi_\mathrm{c}$ for both the $P=1$ and $P=0$ cases. This behavior is due to the fact that the system undergoes BEC and hence a progressively larger macroscopic number of particles occupies the lowest-energy state, $\epsilon=0$, contributing to an abrupt reduction of the average energy and a non-analytic behavior of $E_T(\xi)$. 
In particular, we find that the value of $\xi$ at the critical point is twice as large in the $P=0$ regime than in the $P=1$ case (see Figs.~\ref{fig:res_helike}(c) and (a)).
Additionally, the behaviour of $E_T(\xi)$ is steeper in the polarized case than in the unpolarized case (see Fig.~\ref{fig:exi_bulk_SI} in Appendix \ref{app:pol_vs_unpol}). As a consequence, one has $E^{P=0}_T(\xi=1)>E^{P=1}_T(\xi=1)$ and, conversely, $E^{P=0}_T(\xi=-1)<E^{P=1}_T(\xi=-1)$. The energies of the polarized and unpolarized cases are the same at $\xi=0$, that is $E^{P=0}_T(\xi=0)\equiv E^{P=1}_T(\xi=0)$.

We report in Figs.~\ref{fig:res_helike}(b) and \ref{fig:res_helike}(d) the $\xi_E(T)$ functions, computed numerically from Eqs.~(\ref{eq:gc_N}) and (\ref{eq:gc_E}), at energies corresponding to the critical points of the $E_T(\xi)$ curves (see Figs.~\ref{fig:res_helike}(a) \ref{fig:res_helike}(c) respectively) for $T=4, 6, 8$, and $10$~K.
One can see that these functions are characterized by two regimes: there is a steep increase at low temperatures, until the critical point $\xi_\mathrm{c}$ is reached. Subsequently, the increase is less pronounced. This double regime of $\xi_E(T)$ reflects the fact that the system goes through the superfluid transition. 
The effect of a non-zero polarization is to flatten the $\xi_E(T)$ curves.
Figures~\ref{fig:res_helike}(b) and \ref{fig:res_helike}(d), 
also report a fit the numerical data according to
\begin{equation}\label{eq:csi_exp_mine}
\begin{split}
    \xi_E(T) = & \left( a_0(E) +a_2(E) T^2+a_3(E) T^3 \right) \Theta(T-T_\mathrm{c}) + \\
    &\left(b_0(E) + b_1(E) T + b_2(E) T^2 \right) \bar{\Theta}(T-T_\mathrm{c}), 
\end{split}
\end{equation}
where $T_\mathrm{c}$ is the temperature of the inflection point, $\Theta(T)$ is the Heaviside function and $\bar{\Theta}(T-T_\mathrm{c})=(1-\Theta(T-T_\mathrm{c}))$.
As can be seen, these equations provide a good fit for the $\xi_E(T)$ curves.
Notice that the fitting function of Eq.~(\ref{eq:csi_exp_mine}) has the form of Eq.~(\ref{eq:csi_exp_xiong}) for small $T$, becoming a quadratic polynomial for $T < T_\mathrm{c}$. From the curves of Eq.~(\ref{eq:csi_exp_mine}) one obtains the temperature $T$ where the fermion system attains energy $E$ by solving $\xi_E(T) = -1$.

The fitting function in Eq.~(\ref{eq:csi_exp_mine}) captures the two-regimes observed in the independent-particle model, and emphasizes the role of the BEC transition temperature $T_c$, which always occurs at some $\xi > 0$ in realistic systems independently on the specific form of the interaction. However, in the free-particle case, the BEC transition is hardly apparent in the $\xi_E(T)$ curves (see also appendix \ref{app:free}) and a simple polynomial function is enough to properly describe its shape.~\cite{xiong_2023}. In any case, Eq.~(\ref{eq:csi_exp_mine}) is robust enough to handle all such cases automatically.
In theory, the polynomial expression in Eq.~(\ref{eq:csi_exp_xiong}), with its infinite number of terms, is highly general and, in theory, capable of fitting any complicated function. Nevertheless, in actual calculations where only a limited number of points can be computed, Eq.~(\ref{eq:csi_exp_mine}) provides a more physically grounded expression to guide the extrapolation.

\begin{figure}
   \includegraphics[scale=0.87]{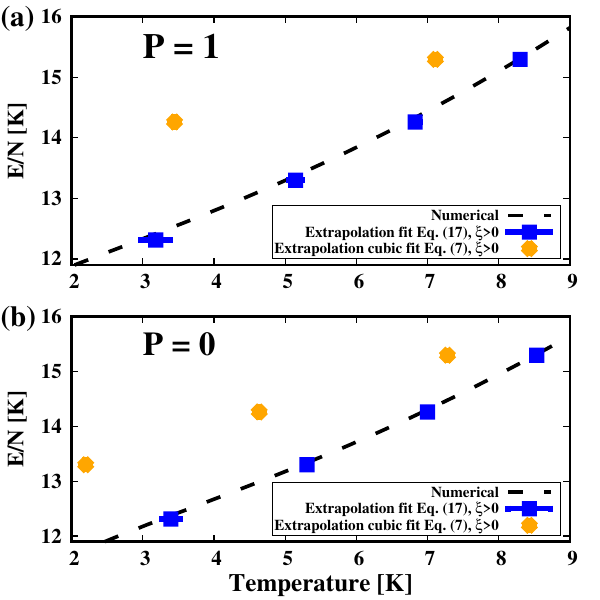}
   \caption{Independent-particle model. (a) Reconstructed energies as function of $T$ for $P=1$. Black dashed line: numerical solution. Blue crosses: extrapolation of Eq.~(\ref{eq:csi_exp_mine}) fitted using only data for $\xi \geq 0$. Orange circles: cubic extrapolation of Eq.~(\ref{eq:csi_exp_xiong}) fitted using only data for $\xi \geq 0$. (b) Same of (a) for the unpolarized ($P=0$) case. We notice that the $P=0$ uncertainties are smaller than those corresponding to $P=1$, and that in both cases the magnitude of the error bar decreases by increasing the temperature.}
   \label{fig:e_reconstr}
\end{figure}

In practice, one can compute $\xi_E(T)$ only for positive $\xi$, hence we re-fitted the parameters of Eq.~(\ref{eq:csi_exp_mine}) using only results in this region. We show in Figs.~\ref{fig:e_reconstr}(a) and \ref{fig:e_reconstr}(b) for the fully polarized and unpolarized cases respectively, the results of the extrapolation to $\xi = -1$ (blue crosses), compared to the actual value obtained by solving Eqs.~(\ref{eq:gc_N}) and (\ref{eq:gc_E}) (black dashed line). The agreement between the extrapolated and actual energies is excellent.. 
However, at lower energies, the critical value of $\xi$ shifts closer to $\xi = 0$, resulting in a larger uncertainty in the extrapolation. 
In the same figures, we also show the result of the extrapolation to $\xi=-1$ using a simple cubic form from Eq.~(\ref{eq:csi_exp_xiong}),~\cite{xiong_2023} fitted from the data at $\xi \geq 0$. In this case, the energy of the fermionic system is not accurately captured, and in some instances, the cubic polynomial fails to intersect the line $\xi = -1$.

\begin{figure*}
   \includegraphics[scale=0.56]{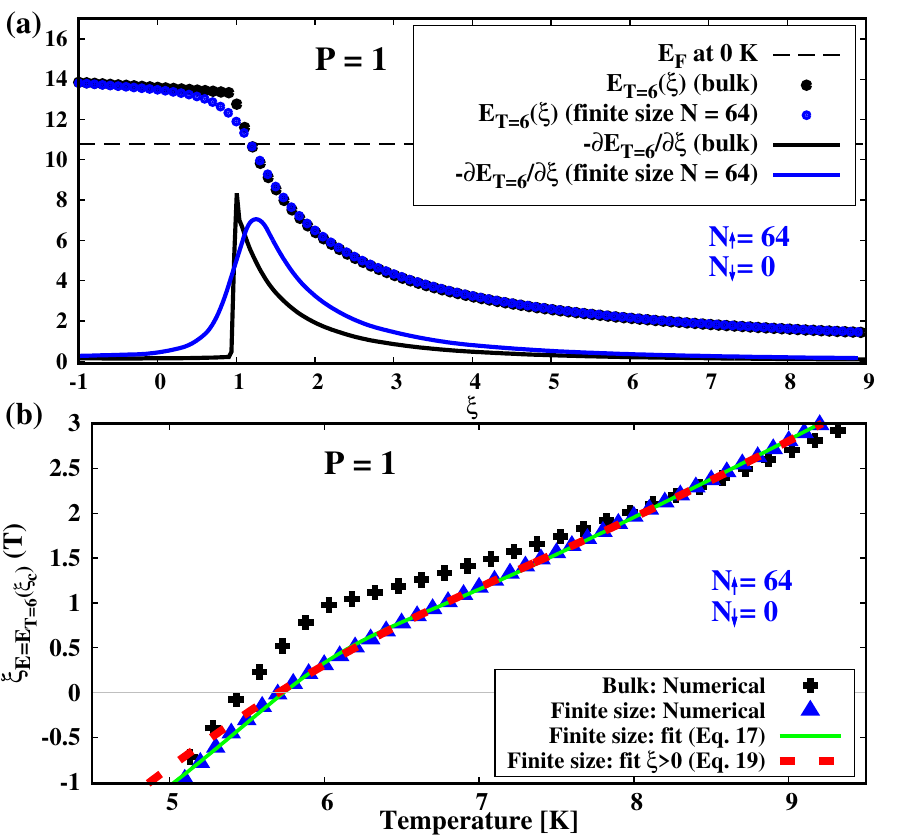}
   \includegraphics[scale=0.56]{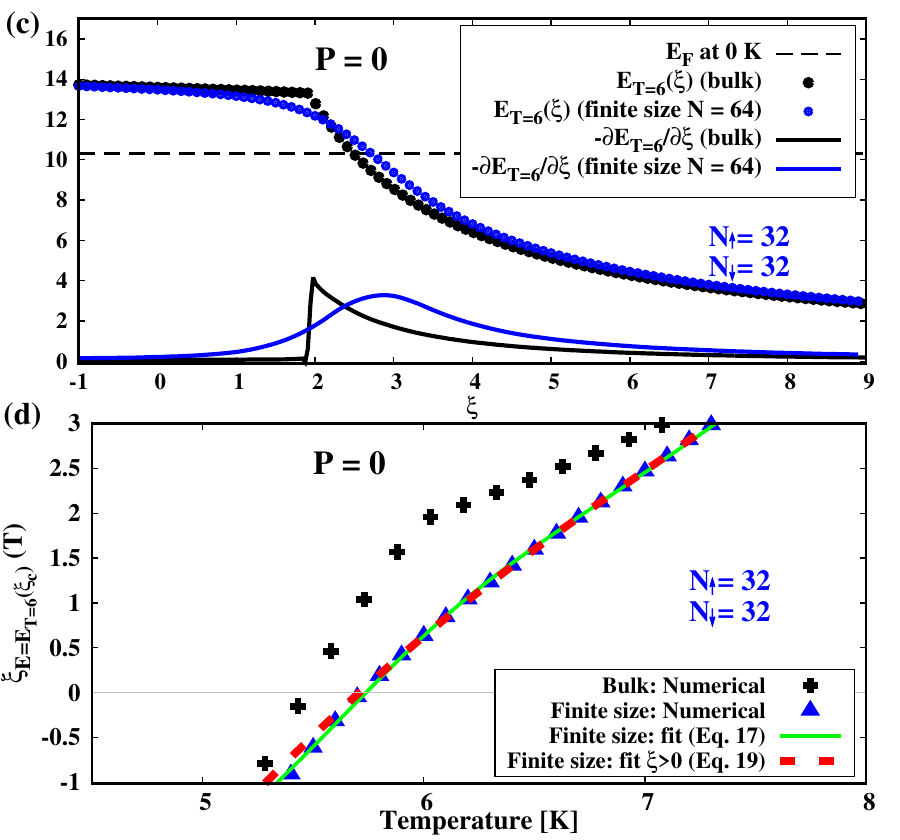}
   \caption{Independent-particle model. (a) Comparison of the energy and its derivative with respect to $\xi$ for $T=6$~K for the bulk system (black points and black short-dashed line) and the finite size system (blue points and blue dashed-line) in the fully polarized case ($P=1$). The long-dashed line corresponds to the Fermi level. (b) Comparison of $\xi_E(T)$ for $E=E_{T=6}(\xi_\mathrm{c})$~K for the bulk system (black circles) versus the finite system (blue triangles). This value for the energy is chosen at the $\xi_\mathrm{c}$ of the bulk system in (a). (c) and (d) Same as (a) and (b) respectively but for $P=0$.}\label{fig:res_bulk_vs_fs}
\end{figure*}
\subsection{Finite-size effects}\label{sec:fs}
In the case of PIMC simulations, the results of energy calculations are generally influenced by finite-size effects. 
In order to study how these limitations would affect the extrapolation to the fermion ($\xi = -1$) case, we can use the recursion formula for $N$ identical particles in the canonical ensemble developed in Ref.~\onlinecite{borrmann_1993} for bosons and fermions, generalized to the case of arbitrary $\xi$ as
\begin{equation}\label{eq:borrmann}
    Z(N,\xi)=\frac{1}{N}\sum_{l=1}^{N} \xi^{l-1} S(l) Z(N-l,\xi),
\end{equation}
%where $N$ is the number of particles %having $\sigma$-spin  
where $S(l)=\sum_{i,j,k} \text{exp}\left[l\beta \epsilon(q_{i},q_{j},q_{k})\right]$.~\cite{hirshberg_2019,hirshberg_2020} For a system composed of particles with different spins, the discrete partition function is straightforwardly generalized as $Z(N,\xi)=\prod_\sigma Z(N_{\sigma,\xi})$. In the following, we will set $N\equiv N_{\uparrow}+N_{\downarrow}=64$, which is the same number of particles used in the PIMC simulations of ${}^3$He.

We show in Fig.~\ref{fig:res_bulk_vs_fs}(a) and \ref{fig:res_bulk_vs_fs}(c) the $E_T(\xi)$ curves for $T=6$~K, in the fully polarized case ($P=1$) and unpolarized ($P=0$) case, respectively. One can see that in a finite-size system the non-analytic behavior across the BEC transition is smoothed, as evidenced by the behavior of the derivative $\partial{E_T}/{\partial \xi}$ (blue lines). %Additionally, the energy per particle in the fermion case is also slightly increased with respect to the bulk value (long-dashed lines).

The finite-size smoothing of the $E_T(\xi)$ curves is mirrored into the $\xi_E(T)$ curves, as can be seen in Figs.~\ref{fig:res_bulk_vs_fs}(b) and \ref{fig:res_bulk_vs_fs}(d) (blue triangles). The inflection due to BEC is much less evident, and finite-size effects generally flatten the $\xi_E(T)$ curve in its proximity. Nevertheless, the energy in the fermion limit ($\xi = -1$) is changed only slightly.
In this case, the critical value of $\xi$ turns out to be very close to 0, preventing a a good fit with the functional form of Eq.~(\ref{eq:csi_exp_mine}). 
Therefore, we found it necessary to slightly change the way of extrapolating results to the fermion case. We notice that the behavior of $\xi_E(T)$ close to $\xi = 0$ is effectively linear, and hence we adopt a fitting function of the form:
\begin{equation}\label{eq:csi_exp_mine_linear}
\begin{split}
    \tilde{\xi}_E(T) = & \left( a_0(E) +a_2(E) T \right) \Theta(T-T_\mathrm{c}) + \\
    &\left(b_0(E) + b_1(E) T +b_2(E)T^2 \right) \bar{\Theta}(T-T_\mathrm{c}), 
\end{split}
\end{equation}
where $T_\mathrm{c}$ is estimated as the temperature after which the non-linear form of $\xi_E(T)$ is apparent.
We stress that the linear term in Eq.~(\ref{eq:csi_exp_mine_linear}) is not meant to describe the physics for $T \to 0$, where $\xi_E$ can take values well below $-1$, but only as a simple and effective extrapolation method to $\xi = -1$.
Notice that the $\xi_E(T)$ curves show an evident linear behavior close to $\xi = 0$ even in the bulk model. 

In Figures~\ref{fig:res_bulk_vs_fs}(b) and \ref{fig:res_bulk_vs_fs}(d) the red dashed line represents the fit using Eq.~(\ref{eq:csi_exp_mine_linear}) taking into account only $\xi_E\geq0$ finite size points, while the green continuous curve is the fit taking into account the points for which $\xi_E\geq-1$. 
Note that the linear extrapolation introduces a slight uncertainty,  $\delta T_\mathrm{fs}\eqsim0.15$~K in the position of the temperature at which $\xi = -1$ is crossed.

\begin{figure*}
   \includegraphics[scale=0.6]{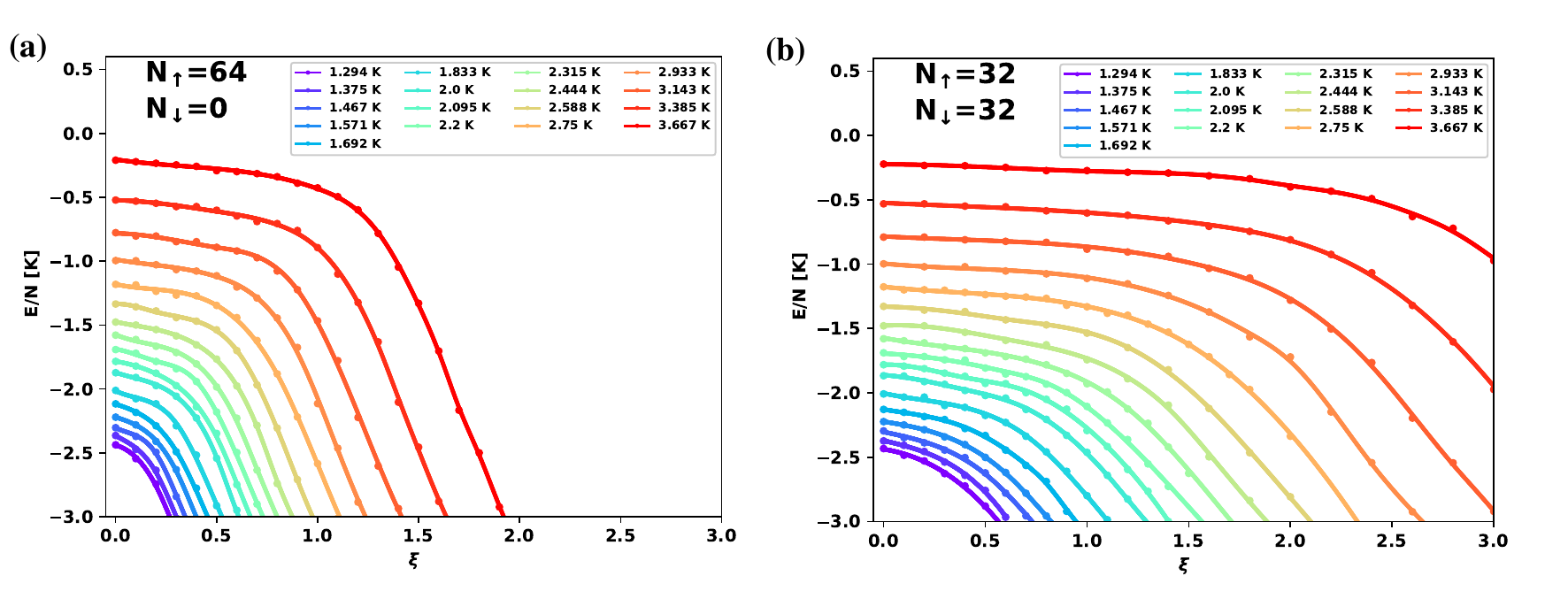}
   \caption{PIMC results for $^3$He. Energy as a function of the $\xi$ parameter for temperatures $T$ in the range of ($1.294$, $3.667$)~K in the fully polarized (panel (a)) and unpolarized (panel (b)) cases. 
   }\label{fig:res_helium3}
\end{figure*}
\begin{figure*}
    \centering
    \includegraphics[width=0.95\linewidth]{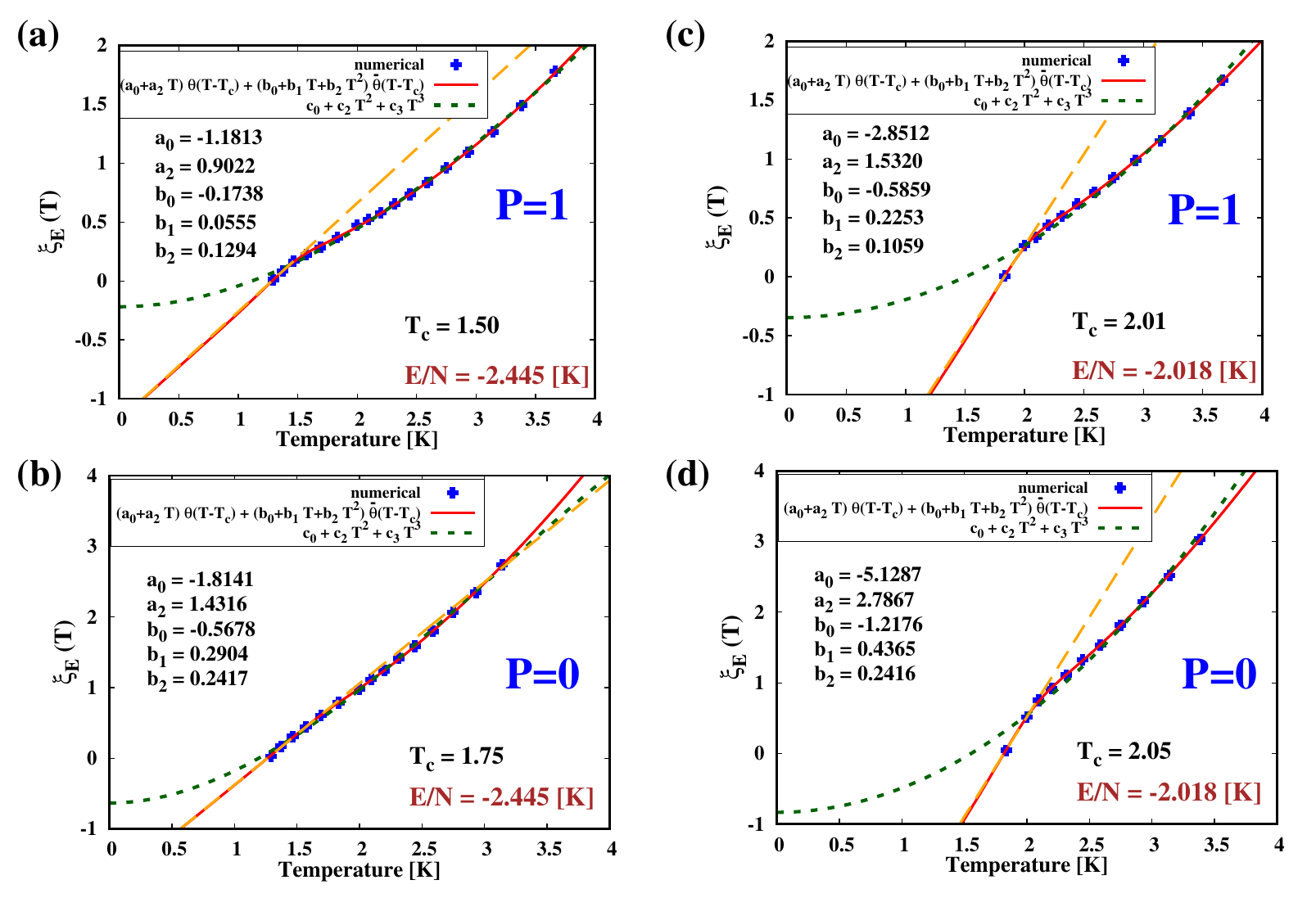}
    \caption{Plot of $\xi_E(T)$ in the $^3$He case from PIMC simulations for two different energies per particle $E/N=-2.445$~K (panels (a) and (b)) and $E/N=-2.018$~K (panels (c) and (d)). Panels (a) and (c) are obtained in the fully polarized case while (b) and (d) in the unpolarized one. We report the values of the fit using Eq.~(\ref{eq:csi_exp_mine_linear}) (red lines), the linear fit passing through the points for $\xi\sim 0$ (orange long-dashed lines) and the cubic fit using Eq.~(\ref{eq:csi_exp_xiong}) (dark-green dashed lines). The value of $T_\mathrm{c}$ has been estimated using the procedure reported in the main text.}
    \label{fig:interp_he3}
\end{figure*}
\section{PIMC results for $^3$He}

In the case of $^3$He simulations, we used the PIMC approach~\cite{ceperley_1986,ceperley_1995} to sample the partition function and the worm algorithm to sample permutations.~\cite{boninsegni_2006} Details of PIMC algorithm are presented elsewhere.~\cite{spada_2022,morresi2025} The only modification to the original set of MC moves,\cite{spada_2022} is the one including the $\xi$ factor of Eq.~(\ref{eq:zxi}) in the weight factor of the \textit{swap} move, the one responsible for the exchange permutation sampling. We considered a system of $N=64$ atoms at a density $\rho=0.016355$~atoms/\AA$^3$, using periodic boundary conditions. We employed the most recent {\em ab-initio} two-body potential to treat the interaction between $^3$He atoms~\cite{czachorowski_2020} and we evaluated the propagator in the pair-product approximation~\cite{pollock_1984} with fixed $\tau=\frac{\beta}{P}=\frac{1}{44}$~K$^{-1}$, where $P$ is the number of beads (imaginary-time slices) of the PIMC approach. For each temperature, $P$ is thus chosen as $P=\text{int} \left[\frac{1}{\tau T} \right]$, where $\mathrm{int}(x)$ is the closest integer number to $x$. Energies are estimated using the virial estimator \cite{ceperley_1995} and then are corrected perturbatively by adding the three-body potential contribution.~\cite{lang_2023}
Results of the $E_T(\xi)$ for $^3$He are shown in Fig.~\ref{fig:res_helium3}. Panel (a) reports the energies obtained as a function of the $\xi$ parameter at temperatures ranging from $1.294$~K to $3.667$~K for $P=1$ while panel (b) the same $E_T(\xi)$ but for $P=0$. The points obtained from PIMC calculations are interpolated using B-splines.~\cite{dierckx_1975} The behaviour of the $E_T(\xi)$ closely resembles the finite-size calculations using the helium-like dispersion of Eq.~(\ref{eq:disp_helium-like}), with a $T$-dependent smoothed inflection point where the superfluid transition occurs. 

Similar to the independent-particle system discussed above, both the unpolarized and polarized systems converge to the same value of the energy for $\xi \to 0$ and the inflection points of $E_T(\xi)$ are shifted towards larger $\xi$-values in the unpolarized case. This is a consequence of the fact that the unpolarized system is "less bosonic" as only half of the particles participate in permutational exchanges. Consequently, the energy remains closer to $E_T(\xi=0)$ for larger values of $\xi$.

\begin{figure}
    \centering
    \includegraphics[width=0.99\linewidth]{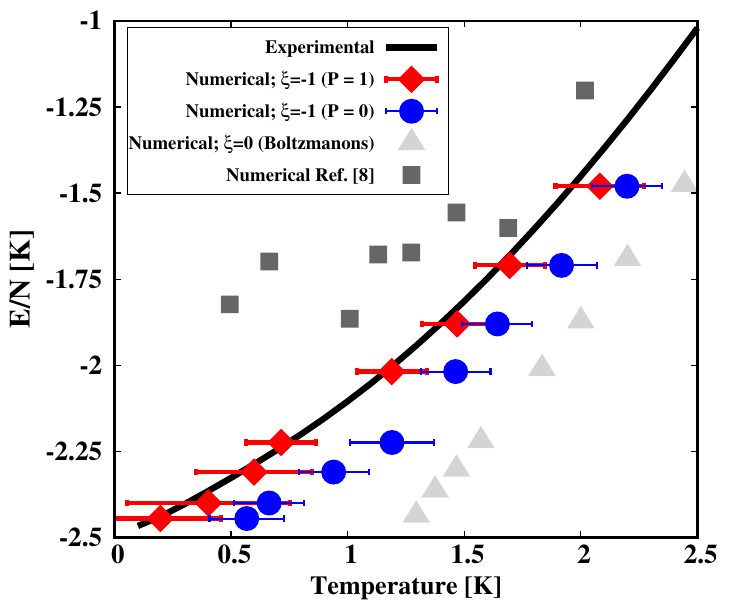}
    \caption{Total energy per particle as a function of temperature from PIMC simulations. Black line corresponds to the experimental data from Ref.~\cite{greywall_1983}; red diamonds are the results extrapolated to $\xi=-1$ in the fully polarized case ($P=1$), blue circles are the one extrapolated to $\xi=-1$ in the unpolarized case ($P=0$), light-gray triangular points are the $\xi=0$ (Boltzmannons) energies while dark-gray square points are the numerical results obtained using PIMC plus fixed-node approximation in Ref.~\cite{ceperley_1992}. Details of error bar computation are provided in the text.}
    \label{fig:ene_he3}
\end{figure}
We show in Fig.~\ref{fig:interp_he3} some $\xi_E(T)$ functions, estimated by cutting the curves in Fig.~\ref{fig:res_helium3}(a) (see Figs.~\ref{fig:interp_he3}(a) and (c)) and in Fig.~\ref{fig:res_helium3}(b) (see Figs.~\ref{fig:interp_he3}(b) and (d)) at constant energies. Blue points represent numerical data, while the red lines are the fits using Eq.~(\ref{eq:csi_exp_mine_linear}). For the sake of reproducibility, the values of the parameters are reported in those figures. The dark-green dashed lines represent fits using a simple cubic polynomial, as used in the original papers on this method.~\cite{xiong_2023} 
Notice that in all the reported cases this fitting function does not intersect the $\xi=-1$ line, which leads to an unphysical result: it would imply that no temperature corresponds to the assumed value of the energy per particle. 
Fitting Eq.~(\ref{eq:csi_exp_mine_linear}) to the numerical data for a given energy $E$, turns our to be more difficult for the unpolarized case (panels (b) and (d)) than for the polarized case since the $T_\mathrm{c}$ is more difficult to estimate from the available data. We estimated it using the following procedure: we fitted the data with a function which is linear below a certain $T_0$ and quadratic above. For each value of $T_0$ we computed the $\chi$-squared of the optimal fit, $\chi^2(T_0)$, and we chose $T_c$ as the position of the minimum of $\chi^2(T_0)$.
We emphasize that this procedure would be much smoother for larger number of particles $N$.

Finally, we show in Fig.~\ref{fig:ene_he3} and Table~\ref{tab:tab_ene} the results for the energy per particle obtained by extrapolation of the fitting function in Eq.~(\ref{eq:csi_exp_mine_linear}). 
We also report an estimate of the uncertainty of our results, coming from the uncertainty in the location of the solution of $\xi_E(T) = -1$. 
This contribution to the uncertainty is larger in the fully polarized case than in the unpolarized one, as in the independent-particle model (see Fig.~\ref{fig:e_reconstr} above) and has been estimated by varying the $T_\mathrm{c}$ in Eq.~(\ref{eq:csi_exp_mine_linear}) by $0.2$~K and re-extrapolating. Specifically, this variation, denoted as $\delta T_\mathrm{c}$, represents the main source of error in determining $E(T)/N$,  being larger than the uncertainties arising from the PIMC simulations and the interpolation of $E_T(\xi)$ data points at $\xi\geq 0$. 
Considering also the systematic uncertainty estimated from the independent-particle model in the finite size case, that is the contribution $\delta T_\mathrm{fs}$  discussed in sec.~\ref{sec:fs}, the error bar $\delta T$ shown in Fig.~\ref{fig:ene_he3} was computed as $\delta T=\sqrt{\delta T_\mathrm{c}^2+\delta T_\mathrm{fs}^2}$.

As expected from the independent particle model and from previous results,~\cite{Bouchaud87} we found that the energy of the spin-unpolarized system is lower than that of the fully polarized one by $\approx 0.2$~K.
The same figure also shows the experimental results from Ref.~\onlinecite{greywall_1983}, which have been obtained by integrating the specific heat curve and then rigidly shifted by $-2.4788$~K~\cite{Bouchaud87} to match the experimental value at $T=0$.
Our results show a fairly good agreement with experimental data considering the uncertainties of our calculation, although we notice a systematic underestimation of the energy per particle by $\approx 0.2$~K.
%However, unexpectedly, the experimental results from Ref.~\onlinecite{greywall_1983}, which has been obtained by integrating the specific heat curve and then rigidly shifted by $-2.4788$~K~\cite{Bouchaud87} to match the experimental value at $T=0$, are in good agreement with the $P=1$ energy (red diamond in Fig.~\ref{fig:ene_he3}) while the $P=0$ points (blue circles) are shifted at lower energies of $\sim 0.2$ K.} 
We also report in Fig.~\ref{fig:ene_he3} the energy per particle without exchange effects (Boltzmannons, that is $\xi=0$, light-grey trinagles) and the results obtained using PIMC in conjunction with a fixed-node constraint (dark-grey squares):~\cite{ceperley_1992} the former strongly underestimate the fermion energies, while the latter are closed to the experimental value, but display a more scattered behavior.

\begin{table}[h!]
\centering
\begin{tabular}{c  ||  c |  c | c } 
 $E/N$~[K]  & $T$~[K], $\xi=-1$ & $T$~[K], $\xi=-1$ &  $T$~[K], $\xi=0$ \\
  & $P=1$ & $P=0$ & \\
 \hline\hline
$-2.445$ & $0.19 \pm 0.30$ & $0.57 \pm 0.22$ &  $1.29$  \\
$-2.400$ & $0.40 \pm 0.38$ & $0.66 \pm 0.21$ &   $1.33$ \\
$-2.310$ & $0.60 \pm 0.29$ & $0.94 \pm 0.21$ & $1.45$ \\
$-2.224$ & $0.71 \pm 0.21$ & $1.19 \pm 0.23$ & $1.56$ \\
$-2.018$ & $1.19 \pm 0.20$& $1.46 \pm 0.19$ & $1.82$ \\
$-1.880$ & $1.47 \pm 0.21$ & $1.64 \pm 0.20$ & $1.99$ \\
$-1.710$ & $1.69 \pm 0.18$ & $1.92 \pm 0.18$ & $2.18$ \\
$-1.480$ & $2.08 \pm 0.24$ & $2.20 \pm 0.17$ & $2.44$ \\
 \hline
\end{tabular}
\caption{Summary of the numerical energies and temperatures reported in Fig.~\ref{fig:ene_he3}. All the energies are given in Kelvin units.}
\label{tab:tab_ene}
\end{table}

\section{Conclusions and perspectives}

We presented simulations of normal liquid ${}^3$He where the fermionic nature of the system was taken into account using a recently developed approach based on a $\xi$-parametrized partition function that continuously interpolates between bosonic ($\xi=1$) and fermionic ($\xi=-1$) statistics.
By studying an independent-particle model where results are numerically exact, we found that the presence of the superfluid transition in the $\xi > 0$ sector, where calculations were not affected by the sign problem, induced significant non-analytic behavior in the function $E(T,\xi)$ that required a tailored approach to extrapolate to the fermion case. 

Our simulations, performed for polarized and unpolarized liquid ${}^3$He, yielded good agreement with experimental data for the energy of particles, although the $P=0$ results underestimate systematically the experimental curve by $\sim 0.2$~K.
While the aim of the work was the extension of the formalism previously developed in Ref.~\onlinecite{xiong_2023}, the approach developed here may find useful application in other strongly-interacting fermionic systems, such as those relevant to nuclear physics.

\section*{Acknowledgements}
TM and GG thank Gabriele Spada and Stefano Giorgini for useful discussions and for the details of the PIMC+worm algorithm. We also extend our gratitude to an anonymous reviewer for their valuable feedback and constructive comments.

\appendix
\section{Independent-particle model: Polarized vs unpolarized energies}\label{app:pol_vs_unpol}
\begin{figure}[h!]
    \centering
    \includegraphics[width=0.99\linewidth]{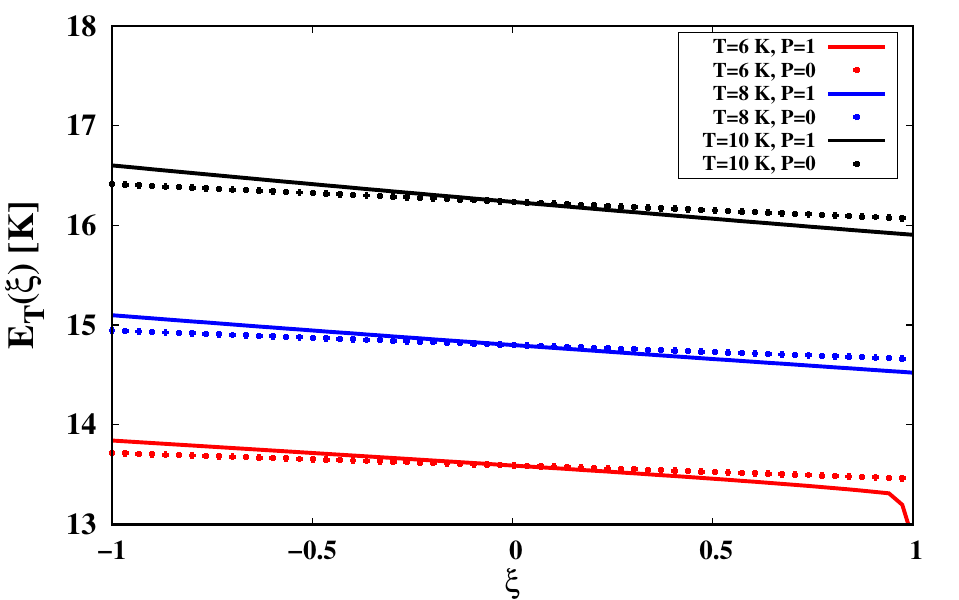}
    \caption{Independent-particle model. Energies as a function of the $\xi$ parameter for the dispersion in Eq.~(\ref{eq:disp_helium-like}) of the main text and for $P=1$ (continuous lines) and $P=0$ (circles) from Eqs.~(\ref{eq:gc_N}) and (\ref{eq:gc_E}) of the manuscript for $T=6$ (red color), $T=8$ (blue color) and $T=10$~K (black color).}
    \label{fig:exi_bulk_SI}
\end{figure}

While the energy of the independent-particle model has been reported in Figs.~\ref{fig:res_helike}(a) and ~\ref{fig:res_helike}(c) for the polarized and unpolarized cases respectively, here we directly compare those energies in figure~\ref{fig:exi_bulk_SI}. 
The direct comparison is useful to highlight the different behaviour of the $P=0$ (points) and $P=1$ (continuous lines) energies in the range of $\xi$ including all the physical values, \textit{i.e.} $\xi=-1,0,1$. From Fig.~\ref{fig:exi_bulk_SI} one can observe the flattening of the energy dispersion around the Boltzmannon energy ($\xi=0$) induced by the condition $P=0$. In the latter case, the system consists of two distinct types of particles with different spins. As a result, for a fixed total density, the system is less fermionic and less bosonic since only half of the particles are involved in permutational exchanges. Therefore, for a spin-independent Hamiltonian, the fermionic system always favours the unpolarized state having lower energy.

\section{Independent-particle model: Free dispersion}\label{app:free}
\begin{figure}
   \includegraphics[scale=0.54]{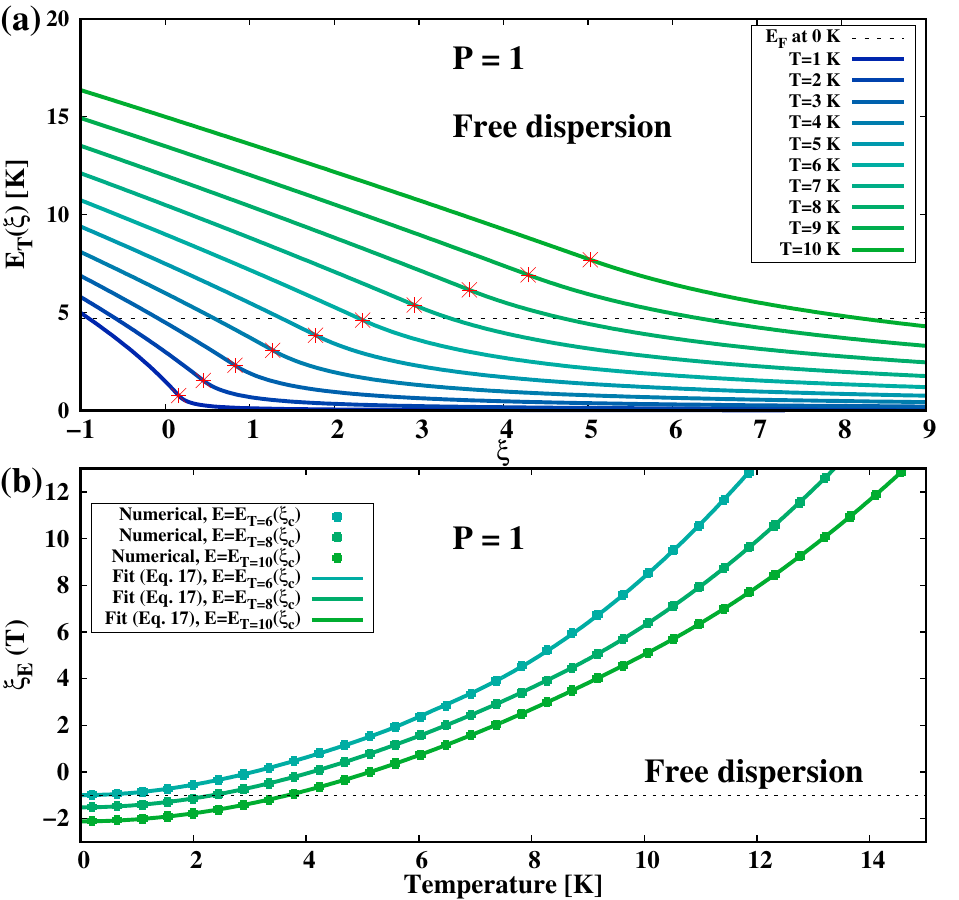}
   \caption{Independent--particle model. (a) Energies as a function of the $\xi$ parameter for the free dispersion (obtained by setting $\alpha_1=0$ in Eq.~(\ref{eq:disp_helium-like}) of the manuscript) and for $P=1$ (fully polarized case) from Eqs.~(\ref{eq:gc_N}) and (\ref{eq:gc_E}) for temperatures between $1$~K and $10$~K. Red stars indicate the energies obtained at $\xi_\mathrm{c}$ in Eq.~(\ref{eq:xicrit}). (b) Behaviour of $\xi_E(T)$ for a few selected energies for $P=1$. The latter are chosen at the critical points for $6, 8$, and $10$~K. Notice the absence of slope change of $\xi_E(T)$ at these temperatures, at variance with Fig.~\ref{fig:res_helike}(b).} 
   \label{fig:res_free}
\end{figure}

By using the same dispersion of Eq.~(\ref{eq:disp_helium-like}) with $\alpha_1=0$, we solved numerically Eqs.~(\ref{eq:gc_N}) and (\ref{eq:gc_E}) fixing again the density at $m N/\Omega=0.081$~g/cm$^3$. Results are shown in Fig.~\ref{fig:res_free} for the fully polarized system. In panel (a) we report the $E_T(\xi)$ dispersions for temperatures ranging from $1$ to $10$~K. Red stars indicate the critical $\xi$ points at which the system undergoes the BEC transition.
We notice that at variance with the behaviour of $E_T(\xi)$ in Fig.~\ref{fig:res_helike}(a), here the inflection point is much smoother due to the different single-particle dispersion. 

This is reflected in panel (b) in the shape of the $\xi_E(T)$, which follows a parabolic behaviour as predicted in Ref.~\cite{xiong_2023}. In this case, Equation~\ref{eq:csi_exp_mine} clearly represents an overfit to $\xi_E(T)$ as the critical temperature is not evident. Nevertheless, Eq.~(\ref{eq:csi_exp_mine}) automatically embodies the possible inflection point that can be driven by superfluid effects in strongly correlated systems.  

\clearpage
\bibliography{main}% Produces the bibliography via BibTeX.

\end{document}